\documentclass[aps,twocolumn,pra,superscriptaddress,showpacs,tightenlines]{revtex4-1}
\usepackage{amssymb}
\usepackage{amsmath}
\usepackage{graphicx}
\usepackage{epsfig}
\usepackage{subfigure}
\usepackage{amsfonts}
\usepackage{txfonts}
\usepackage{CJK}
\usepackage[colorlinks,citecolor=blue, linkcolor=blue,hyperindex,CJKbookmarks,dvipdfm]{hyperref}
\begin{document}

%\begin{CJK*}{GBK}{song}
\title{Photon blockade induced by atoms with Rydberg coupling}
\author{Jin-Feng Huang }
\affiliation{State Key Laboratory of Theoretical Physics, Institute of Theoretical Physics,
Chinese Academy of Sciences, and University of the Chinese
Academy of Sciences, Beijing 100190, China}
\author{Jie-Qiao Liao }
\affiliation{Department of Physics and Institute of Theoretical Physics,
The Chinese University of Hong Kong, Shatin, Hong Kong SAR, China}
\affiliation{Advanced Science Institute, RIKEN, Wako-shi, Saitama 351-0918, Japan}
\author{C. P. Sun }
\email{suncp@itp.ac.cn}
\homepage{http://power.itp.ac.cn/ suncp/index.html}
\affiliation{State Key Laboratory of Theoretical Physics, Institute of Theoretical Physics,
Chinese Academy of Sciences, Beijing 100190, China}
\affiliation{Beijing Computational Science Research Center, Beijing 100084, China }

\date{\today}

\begin{abstract}
We study the photon blockade of two-photon scattering in a one-dimensional waveguide,
which contains two atoms coupled via the Rydberg interaction. We obtain the analytic scattering solution of photonic
wave packets with the Laplace transform method. We examine the photon correlation by addressing
the two-photon relative wave function and the second-order correlation function in the single- and two-photon
resonance cases. It is found that, under the single-photon resonance condition, photon bunching and antibunching
can be observed in the two-photon transmission and reflection, respectively. In particular, the bunching and
antibunching effects become stronger with the increasing of the Rydberg coupling strength. In addition, we find
a phenomenon of bunching-antibunching transition caused by the two-photon resonance.
\end{abstract}

\maketitle

%文章打成单列的命令%
%%%%%%%%%%%%%%%%%%%%%%%%%%%%%%%%%%%%%%%%%%%%%%%%%%%%%%%%%%%%%%%
%\global\columnwidth20.5pc
%\global\hsize\columnwidth\global\linewidth\columnwidth
%\global\displaywidth\columnwidth
%%%%%%%%%%%%%%%%%%%%%%%%%%%%%%%%%%%%%%%%%%%%%%%%%%%%%%%%%%%%%%%%%%

\section{Introduction\label{sec-intro}}

Controllable transport of photons in one-dimensional (1D) waveguides
is crucial for realizing all-optical quantum devices, which are basic
elements for implementation of photonic quantum information processing.
As an example, a single-photon transistor~\cite{Lukin2007}, controlling the propagation
of a signal photon by another photon, can
be used to implement a single-photon quantum switch. Up to now, there
have been some proposals for coherent transport of single photons and two photons
in waveguides with linear and nonlinear dispersion relations~\cite{Fan2005,Fan2007,Zhou2008,Law2008,Liao2009,
Gong2008,Chang2011,Shi2009,Shi_LSZ,Tsoi2009,Liao2010A,
Liao2010B,Shi2012,Rephaeli2011,Xu2010,Roy2010,Baranger2012}. Most
of them are based on the physical mechanism of photon scattering by
tunable targets.

Two-photon transport in a waveguide is a nontrivial mission since
nonlinear photon-photon interaction~\cite{Imamoglu1997,Kimble2005}
may be induced by nonlinear scattering targets such as a Kerr-type nonlinear
cavity~\cite{Liao2010B} or strongly interacting atoms~\cite{Vuletic2012}.
In particular, there exists a kind of induced photon nonlinearity that can
lead to photon blockade~\cite{Imamoglu1997,Kimble2005}, which can
be used to realize single-photon sources. This photon nonlinearity
in a cavity containing atoms is a second-order effect of the interactions
between the cavity field and the atoms therein~\cite{Imamoglu1997}, thus it is quite
natural to ask how could the atoms directly mediate such a nonlinearity for
photon blockade. Recently, Shi \textit{et al.}~\cite{Shi2011} have considered
two-photon scattering in a waveguide coupled to a cavity containing
a two-level atom. Their results on photon blockade can well fit the
experiment data~\cite{Kimble2005}.

We note that the blockade effect can also occur among atoms. In atomic
blockade, double excitation of atoms is strongly suppressed by some
direct~\cite{Lukin2000,Lukin2001,Tong2004,Saffman2009,Grangier2009} and indirect~\cite{Huang2012} interatom coupling.
Physically, when the interaction between excited atoms is strong enough,
it will be difficult to simultaneously excite two atoms due to the
excess coupling-energy requirement. Since the simultaneous
excitation of two atoms needs to absorb two photons, the atomic blockade
will suppress the simultaneous two-photon absorption. Then the absorbed photons can only be emitted one by one. In other
words, the atomic blockade can induce the photon blockade under certain
conditions. With this motivation, we will investigate in this paper
how the atomic blockade~\cite{Lukin2001,Lukin2001,Tong2004,Saffman2009,Grangier2009,Adams2010,Buchler2011,Lukin2011,Molmer2012,Huang2012} exerts an influence
on the photon blockade~\cite{Imamoglu1997,Kimble2005}.

Deeply understanding the relation between these two kinds of blockade effect
(atomic blockade and photon blockade) will provide a straightforward
way to probe the nature of the inter-atom coupling through photon
blockade effect. It is also possible to display the atomic blockade effect
due to the photon blockade phenomenon. To illustrate the physical
mechanism behind these scientific and technical issues, we study the spatial-wave-packet transport of two photons in a 1D linear waveguide, which contains two
intercoupled two-level Rydberg atoms. To understand the single-photon
contributions in the two-photon process, we first calculate spatial
evolution of the single-photon wave packet. The similar
approach has been used in Refs.~\cite{Liao2010B,Tsoi2009} to study the scattering problem for spatial wave packets. Then, we calculate the evolution of
the spatial wave packet for two photons. Since the second-order
correlation function $g^{(2)}$ can describe the photon-blockade effect,
we calculate $g^{(2)}$ for both the reflected and transmitted photons. We find that the $g^{(2)}$
is proportional to the relative spatial distribution probability of the two photons with some
distance. We can predict the photon statistical properties from the wave
function of the two photons. Our calculation shows that the spatial distribution
of the two photons is strongly dependent on the interatom coupling strength, where we have employed the direct Rydberg coupling between the two atoms.
The transmitted photons are strongly bunched except for some special
Rydberg coupling strength, which depends on the initial conditions. The
reflected photons are strongly repulsed by each other, namely the photon-blockade effect, when Rydberg coupling strength is away from
the two-photon resonance, which means the Rydberg coupling strength equals the sum
of each photon's center detuning. Here, the center detuning denotes the deviation of the frequency center for single input wave packet
from the atomic transition frequency. On the contrary, at this two-photon resonance,
the two reflected photons exhibits bunching behavior. We can use this
sole bunching behavior for the reflected two photons to probe the
nature of the interatom coupling.

This paper is organized as follows. In Sec.~\ref{sec-setup}, we
introduce the physical model and its Hamiltonian. In Secs.~\ref{sec-sinphosca}
and~\ref{sec-twophosca}, we solve the single- and two-photon
scattering problem with the Laplace transform, respectively. In particular,
we describe the two-photon correlation in the two-photon relative
coordinate space, and calculate the second-order correlation function
for the two reflected or transmitted photons. Finally, we draw our conclusion
in Sec.~\ref{sec-conclu} and present the detailed derivations for the
two-photon solution in the Appendix.

\section{Model setup\label{sec-setup}}

\begin{figure}
\includegraphics[bb=45 644 389 766,width=3.3in]{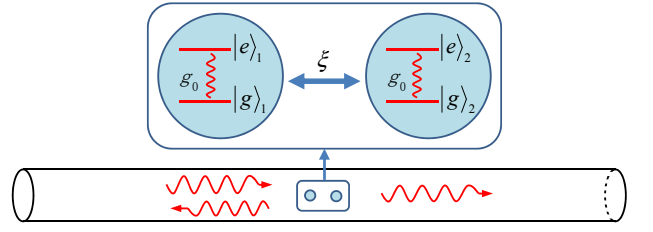} \caption{Schematic diagram of the physical setup. Two interacting atoms are
placed in a 1D linear waveguide. Photons injected from the left-hand
side of the waveguide are scattered by the atoms.}
\label{setup}
\end{figure}

We start by considering a $1$D linear waveguide, which contains two
two-level atoms coupled via the Rydberg interaction (see Fig.~\ref{setup}).
The model Hamiltonian (with $\hbar=1$) of the system reads as
\begin{eqnarray}
\hat{H} & = & \int_{0}^{\infty}dk\omega_{k}(\hat{r}_{k}^{\dagger}\hat{r}_{k}+\hat{l}_{k}^{\dagger}\hat{l}_{k})+\frac{\omega_{0}}{2}(\hat{\sigma}_{1}^{z}+\hat{\sigma}_{2}^{z})\nonumber \\
 &  & +g_{0}\int_{0}^{\infty}dk\sum_{\ell=1,2}[\hat{\sigma}_{\ell}^{+}(\hat{r}_{k}+\hat{l}_{k})+(\hat{r}_{k}^{\dagger}+\hat{l}_{k}^{\dagger})\hat{\sigma}_{\ell}^{-}]\nonumber \\
 &  & +\xi\vert e\rangle_{1}\langle e\vert_{1}\otimes\vert e\rangle_{2}\langle e\vert_{2}.\label{Horiginal}
\end{eqnarray}
The first line of Eq.~(\ref{Horiginal}) is the free Hamiltonian
of the fields and atoms. The creation (annihilation) operators $\hat{r}_{k}^{\dagger}$
($\hat{r}_{k}$) and $\hat{l}_{k}^{\dagger}$ ($\hat{l}_{k}$) describe,
respectively, the right- and left-propagating light fields in the waveguide,
with wave vector $k$ and frequency $\omega_{k}=\upsilon_{p}k$ (hereafter
we take the group velocity of light $\upsilon_{p}=1$). Pauli's operators $\hat{\sigma}_{\ell}^{x,y,z}$
{[}$\hat{\sigma}_{\ell}^{\pm}=\frac{1}{2}(\hat{\sigma}_{\ell}^{x}\pm i\hat{\sigma}_{\ell}^{y})${]}
are introduced to represent the $\ell$th ($\ell=1,2$) atom with
energy level spacing $\omega_{0}$. The Hamiltonian in the second line
of Eq.~(\ref{Horiginal}) depicts the atom-field interaction with
the coupling strength $g_{0}$. In addition, the last term in Eq.~(\ref{Horiginal})
stands for the Rydberg interaction of strength $\xi$ between the
Rydberg states $|e\rangle_{\ell}$ ($\ell=1,2$) of each atom~\cite{Lukin2001,Adams2010,Cote2002,Marinescu1997,Weidemuller2010,Grangier2012,Wu2011}.
Physically, this Rydberg coupling strength $\xi$ depends on the distance $r$ between the two atoms.
For the dipole-dipole and van der Waals interactions, the strength $\xi$ takes the form $\xi\sim1/r^{3}$ and $\xi\sim1/r^{6}$, respectively.
When the distance is on the order of a few $\mu$m, $\xi$ could be very strong~\cite{Grangier2012}.
Since the coordinates of the two atoms are external parameters rather than dynamical variables, we use the strength $\xi$ to characterize the Rydberg interaction.

By introducing the even- and odd-parity modes,
\begin{equation}
\hat{b}_{k}=\frac{1}{\sqrt{2}}(\hat{r}_{k}+\hat{l}_{k}),\hspace{1cm}\hat{c}_{k}=\frac{1}{\sqrt{2}}(\hat{r}_{k}-\hat{l}_{k}),
\end{equation}
the Hamiltonian $\hat{H}=\hat{H}^{(o)}+\hat{H}^{(e)}$ can be decomposed into two parts,
the odd-parity part $\hat{H}^{(o)}=\int_{0}^{\infty}dk\omega_{k}\hat{c}_{k}^{\dagger}\hat{c}_{k}$
and the even-parity part,
\begin{eqnarray}
\hat{H}^{(e)} & = & \int_{0}^{\infty}dk\omega_{k}\hat{b}_{k}^{\dagger}\hat{b}_{k}+\frac{\omega_{0}}{2}(\hat{\sigma}_{1}^{z}+\hat{\sigma}_{2}^{z})\nonumber \\
 &  & +g\int_{0}^{\infty}dk[(\hat{\sigma}_{1}^{+}+\hat{\sigma}_{2}^{+})\hat{b}_{k}+\hat{b}_{k}^{\dagger}(\hat{\sigma}_{1}^{-}+\hat{\sigma}_{2}^{-})]\nonumber \\
 &  & +\xi\vert e\rangle_{1}\langle e\vert_{1}\otimes\vert e\rangle_{2}\langle e\vert_{2},
\end{eqnarray}
 where $g=\sqrt{2}g_{0}$. Obviously, the odd-parity modes decouple
with the atoms so that their evolution is free. In the following we
will mainly deal with the evolution of the even-parity modes.

In a rotating frame with respect to
\begin{equation}
\hat{H}_{0}^{(e)}=\omega_{0}\int_{0}^{\infty}dk\hat{b}_{k}^{\dagger}\hat{b}_{k}+\frac{\omega_{0}}{2}(\hat{\sigma}_{1}^{z}+\hat{\sigma}_{2}^{z}),
\end{equation}
 the Hamiltonian $\hat{H}^{(e)}$ becomes
\begin{eqnarray}
\hat{H}_{I}^{(e)} & = & \int_{0}^{\infty}dk\Delta_{k}\hat{b}_{k}^{\dagger}\hat{b}_{k}+\xi\vert e\rangle_{1}\langle e\vert_{1}\otimes\vert e\rangle_{2}\langle e\vert_{2}\nonumber \\
 &  & +g\int_{0}^{\infty}dk[(\hat{\sigma}_{1}^{+}+\hat{\sigma}_{2}^{+})\hat{b}_{k}+\hat{b}_{k}^{\dagger}(\hat{\sigma}_{1}^{-}+\hat{\sigma}_{2}^{-})],\label{Hie}
\end{eqnarray}
 where $\Delta_{k}=\omega_{k}-\omega_{0}$. Based on Hamiltonian (\ref{Hie}),
we can work out the photon scattering solution for the system. To
understand the physical picture for the photon scattering,
we will first consider the single-photon scattering, and then we
will study how the Rydberg interaction affects the two-photon
scattering.

\section{Single-photon scattering\label{sec-sinphosca}}

In the following discussions, we will employ the dynamical approach~\cite{Tsoi2009,Liao2010B} rather than the stationary-state approach~\cite{Gong2008,Shi2009,Zhou2008,Chang2011,Xu2010,Fan2007,Fan2005}
to study the photon-scattering problem. We note that the total-excitation-number operator $\hat{N}=\hat{a}^{\dagger}\hat{a}+\vert e\rangle_{1}\langle e\vert_{1}+\vert e\rangle_{2}\langle e\vert_{2}$
in this system is a conserved quantity due to $[\hat{N},\hat{H}_{I}^{(e)}]=0$.
Consequently, the Hilbert space of the system can be divided into
the direct-sum subspaces with different excitations. For single-photon
scattering, it is sufficient to consider the scattering problems within
the single-excitation subspace. In this subspace, an arbitrary state
of the system can be expressed as
\begin{eqnarray}
|\Phi(t)\rangle & = & \alpha_{1}(t)|\emptyset\rangle|e\rangle_{1}|g\rangle_{2}+\alpha_{2}(t)|\emptyset\rangle|g\rangle_{1}|e\rangle_{2}\nonumber \\
 &  & +\int_{0}^{\infty}dk\beta_{k}(t)\hat{b}_{k}^{\dagger}|\emptyset\rangle|g\rangle_{1}|g\rangle_{2},
\end{eqnarray}
 where $|\emptyset\rangle$ is the vacuum state and $|1_{k}\rangle$
represents the state with a single photon in the $k$th even mode
of the waveguide. The variables $\alpha_{1}(t)$, $\alpha_{2}(t)$,
and $\beta_{k}(t)$ are probability amplitudes. By solving the Schr\"{o}dinger
equation $i\left(\partial/\partial t\right)|\Phi(t)\rangle=\hat{H}_{I}^{(e)}|\Phi(t)\rangle$,
we obtain the following equations of motion for probability amplitudes:
\begin{eqnarray}
\dot{\alpha}_{1}(t) & = & -ig\int_{0}^{\infty}dk\beta_{k}(t),\nonumber \\
\dot{\alpha}_{2}(t) & = & -ig\int_{0}^{\infty}dk\beta_{k}(t),\nonumber \\
\dot{\beta}_{k}(t) & = & -i\Delta_{k}\beta_{k}(t)-ig[\alpha_{1}(t)+\alpha_{2}(t)].
\end{eqnarray}

For single-photon scattering, we assume that the two atoms are initially
in ground state and a single even-mode photon is in a Lorentzian wave
packet. The initial condition reads
\begin{eqnarray}
\alpha_{1}(0) & = & 0,\hspace{0.8cm}\alpha_{2}(0)=0,\nonumber \\
\beta_{k}(0) & = & G_{1}\frac{e^{i\Delta_{k}l}}{\Delta_{k}-\delta+i\epsilon},\hspace{0.8cm}l\geq0,
\end{eqnarray}
where $G_{1}=\sqrt{\epsilon/\pi}$ is the normalization constant,
and $l$ is the initial distance between the wavefront of the photon
wave packet and the atoms. In addition, $\delta$ and $\epsilon$ are, respectively, the center
detuning and width of the wave packet. The transient solution of these probability
amplitudes might be obtained using the Laplace transform method. For the scattering problem, we focus on the long-time solution,
\begin{eqnarray}
\alpha_{1}(t\rightarrow\infty) & = & 0,\hspace{0.8cm}\alpha_{2}(t\rightarrow\infty)=0,\nonumber \\
\beta_{k}(t\rightarrow\infty) & = & \bar{t}_{k}\beta_{k}(0)e^{-i\Delta_{k}t}, \label{eq:sol_s}
\end{eqnarray}
where we introduce the phase factor
\begin{equation}
\bar{t}_{k}=\frac{\Delta_{k}-i\gamma}{\Delta_{k}+i\gamma}\label{eq:tbar}
\end{equation}
with $\gamma=2\pi g^{2}$ being the decay rate of each atom. It follows from Eq.~(\ref{eq:sol_s}) that the single photon with wave vector $k$ only picks up a phase shift $\varphi_{k}$ defined by $\exp{(i\varphi_{k})}=\bar{t}_{k}$ after being scattered by the two atoms.

In practice, a single photon should be injected in right- or left-propagating
modes. Hence, we need to consider the single-photon scattering in modes
$\hat{r}_{k}$ and $\hat{l}_{k}$. For the right-propagating single-photon injection,
the initial state is
\begin{eqnarray}
\vert\psi(0)\rangle & = & \int_{0}^{\infty}dk\beta_{k}(0)\hat{r}_{k}^{\dagger}\vert\emptyset\rangle\nonumber \\
 & = & \frac{1}{\sqrt{2}}\int_{0}^{\infty}dk\beta_{k}(0)(\hat{b}_{k}^{\dagger}+\hat{c}_{k}^{\dagger})\vert\emptyset\rangle.
\end{eqnarray}
The injected single-photon wave function in position space reads
\begin{equation}
\langle x\vert\psi(0)\rangle\simeq-i\sqrt{2\pi}G_{1}e^{i(\delta-i\epsilon)(x+l)}\theta(-x-l),
\label{eq:psi0}
\end{equation}
where $\theta(x)$ is the Heaviside step function. In the long-time
limit, the single-photon wave function in $k$ space is
\begin{eqnarray}
\vert\psi(t\rightarrow\infty)\rangle & = & \int_{0}^{\infty}dk\beta_{k}(0)e^{-i\Delta_{k}t}(T_{k}\hat{r}_{k}^{\dagger}+R_{k}\hat{l}_{k}^{\dagger})\vert\emptyset\rangle,\nonumber \\
\end{eqnarray}
where the transmission and reflection amplitudes are obtained as
\begin{eqnarray}
T_{k}=\frac{\Delta_{k}}{\Delta_{k}+i\gamma},\hspace{0.8cm}R_{k}=\frac{-i\gamma}{\Delta_{k}+i\gamma}.
\end{eqnarray}
The above result has been reported in Refs.~\cite{Fan2005,Liao2010B}. At the resonant case, i.e., $\Delta_{k}=0$, we get $T_{k}=0$ and $R_{k}=-1$, which means that the single
photon is completely reflected. This complete reflection can be explained based on quantum interference.
In the resonant case, the coherent amplitude for transmitted (reflected) photons disappears (increases) due to destructive (constructive) interference between the photon injection and the atom emission channels.

The output wave function in position space is
\begin{eqnarray}
\langle x\vert\psi(t\rightarrow\infty)\rangle & = & \psi_{r}(x,l,\delta)+\psi_{l}(x,l,\delta),\label{eq:output_x_1}
\end{eqnarray}
where
\begin{eqnarray}
\psi_{r}(x,l,\delta) & = & -i\sqrt{2\pi}G_{1}T_{\delta-i\epsilon}e^{i(\delta-i\epsilon)(x-t+l)}\theta(-x+t-l),\nonumber \\
\psi_{l}(x,l,\delta) & = & -i\sqrt{2\pi}G_{1}R_{\delta-i\epsilon}e^{-i(\delta-i\epsilon)(x+t-l)}\theta(x+t-l).\nonumber \\
\label{sinphowaveform}
\end{eqnarray}
We note that the output waveforms (proportional to $|\psi_{r}(x,l,\delta)|^{2}$ and $|\psi_{l}(x,l,\delta)|^{2}$) of the transmitted and reflected photons
are the same as that of the input state (\ref{eq:psi0}), except for their normalized amplitudes by the transmission and reflection coefficients ($|T_{\delta-i\epsilon}|^{2}$ and $|R_{\delta-i\epsilon}|^{2}$), respectively.

\section{Two-photon scattering\label{sec-twophosca}}

We now turn to the two-photon scattering. We will use the Laplace
transform to obtain the long-time state of two photons,
and special attention will be paid to clarifying the relationship
between the atom blockade and the photon blockade.

\subsection{Equations of motion and solution}

In the two-excitation subspace, there are four types of basis state: two excitations in atoms (with
basis $|\emptyset\rangle|e\rangle_{1}|e\rangle_{2}$), one excitation in atoms
and the other in light fields (with bases $|1_{k}\rangle|e\rangle_{1}|g\rangle_{2}$
and $|1_{k}\rangle|g\rangle_{1}|e\rangle_{2}$), and two excitations in light fields (with
basis $|1_{p},1_{q}\rangle|g\rangle_{1}|g\rangle_{2}$). Then an arbitrary
state in this subspace is written as
\begin{eqnarray}
|\varphi(t)\rangle & = & A(t)|\emptyset\rangle|e\rangle_{1}|e\rangle_{2}+\int_{0}^{\infty}dkB_{k}(t)\hat{b}^\dagger_{k}|\emptyset\rangle|e\rangle_{1}|g\rangle_{2}\nonumber \\
 &  & +\int_{0}^{\infty}dkC_{k}(t)\hat{b}^\dagger_{k}|\emptyset\rangle|g\rangle_{1}|e\rangle_{2}\nonumber \\
 &  & +\int_{0}^{\infty}dp\int_{0}^{p}dqD_{p,q}(t)\hat{b}^\dagger_{p}\hat{b}^\dagger_{q}|\emptyset\rangle|g\rangle_{1}|g\rangle_{2},
\end{eqnarray}
where $A(t)$, $B_{k}(t)$, $C_{k}(t)$, and $D_{p,q}(t)$ are probability
amplitudes. By solving the Schr\"{o}dinger equation, we get the following equations
of motion for these probability amplitudes:
\begin{eqnarray}
\dot{A}(t) & = & -i\xi A(t)-ig\int_{0}^{\infty}dk[B_{k}(t)+C_{k}(t)],\nonumber \\
\dot{B}_{k}(t) & = & -i\Delta_{k}B_{k}(t)-igA(t)-ig\int_{0}^{\infty}dpD_{p,k}(t),\nonumber \\
\dot{C}_{k}(t) & = & -i\Delta_{k}C_{k}(t)-igA(t)-ig\int_{0}^{\infty}dpD_{p,k}(t),\nonumber \\
\dot{D}_{p,q}(t) & = & -i(\Delta_{p}+\Delta_{q})D_{p,q}(t)-ig[B_{p}(t)+B_{q}(t)]\nonumber \\
 &  & -ig[C_{p}(t)+C_{q}(t)],\label{eq:motion_eq}
\end{eqnarray}
where we have used the symmetry relation $D_{k,p}(t)=D_{p,k}(t)$.

We consider the initial state where the two atoms are in their ground
state and the two photons are in a wave packet. The corresponding
initial condition reads
\begin{eqnarray}
A(0) & = & 0,\hspace{0.8 cm}B_{k}(0)=0,\hspace{0.8 cm}C_{k}(0)=0,\nonumber\\
D_{p,q}(0) & = & G_{2}\left(\frac{e^{i\Delta_{p}l_{1}}}{\Delta_{p}-\delta_{1}+i\epsilon}\frac{e^{i\Delta_{q}l_{2}}}{\Delta_{q}-\delta_{2}+i\epsilon}\right.\nonumber \\
 &  & \left.+\frac{e^{i\Delta_{q}l_{1}}}{\Delta_{q}-\delta_{1}+i\epsilon}\frac{e^{i\Delta_{p}l_{2}}}{\Delta_{p}-\delta_{2}+i\epsilon}\right),\label{eq:Ini}
\end{eqnarray}
where $l_{\ell}$ is the initial position of the $\ell$th \textbf{($\ell=1,2$)} photon.
Without loss of generality, we assume $l_{1}\geq0$, $l_{2}\geq0$,
and $l_{1}\geq l_{2}$. The normalization constant reads
\begin{equation}
G_{2}=\frac{\epsilon}{\pi}\left[1+\frac{4\epsilon^{2}e^{-2\epsilon\left(l_{1}-l_{2}\right)}}{\left(\delta_{1}-\delta_{2}\right)^{2}+4\epsilon^{2}}\right]^{-1/2}.
\end{equation}

Under the above initial condition, the long-time solution for these
probability amplitudes can be obtained as $A(t\rightarrow\infty)=0$, $B_{k}(t\rightarrow\infty)=0$,
$C_{k}(t\rightarrow\infty)=0$, and
\begin{eqnarray}
D_{p,q}(t\rightarrow\infty) & = & [\bar{t}_{p}\bar{t}_{q}D_{p,q}(0)+J_{p,q}]e^{-i(\Delta_{p}+\Delta_{q})t},\label{eq:Dpq}
\end{eqnarray}
where $\bar{t}_{p}$ $(\bar{t}_{q})$ has been defined by Eq. (\ref{eq:tbar}) and
\begin{eqnarray}
J_{p,q}&=&4G_{2}\gamma^{2}\frac{e^{i(\Delta_{p}+\Delta_{q})l_{1}}}{(\Delta_{p}+i\gamma)(\Delta_{q}+i\gamma)}
\frac{(\Delta_{p}+\Delta_{q}-2\xi)}{(\Delta_{p}+\Delta_{q}-\xi+i\gamma)}\nonumber\\
&&\times\frac{1}{(\Delta_{p}+\Delta_{q}-\delta_{1}-\delta_{2}+2i\epsilon)}\nonumber \\&  & \times\left[\left(\frac{e^{-(i\delta_{2}+\epsilon)(l_{1}-l_{2})}}{i\gamma+\delta_{2}-i\epsilon}
+\frac{e^{-(i\delta_{2}+\epsilon)(l_{1}-l_{2})}}{\Delta_{p}+\Delta_{q}-\delta_{2}+i\epsilon+i\gamma}\right)\right.\nonumber \\
&  & \left.-\left(\frac{e^{-\gamma(l_{1}-l_{2})}}{i\gamma+\delta_{2}-i\epsilon}-\frac{e^{-\gamma(l_{1}-l_{2})}}{\Delta_{p}+\Delta_{q}-\delta_{1}+i\epsilon+i\gamma}\right)\right].
\end{eqnarray}
The first term in Eq. (\ref{eq:Dpq}) describes two-photon
independent scattering process, while the second term represents photon
correlation induced by scattering process. It should be pointed out that,
when $\xi=0$, $J_{p,q}\neq0$. This fact means that the photon correlation
can be observed even in the absence of the Rydberg interaction. Physically,
this residual photon correlation is generated due to the quantum interference
between the two transition channels of the two atoms. A similar result has been shown in Ref.~\cite{Rephaeli2011}.

\subsection{Two-photon wave functions in real space}

We consider a realistic case with the two photons injected
from the left-hand side of the waveguide. Then, the initial state of
the photons can be written as
\begin{eqnarray}
\vert\psi(0)\rangle & = & \int_{0}^{\infty}dp\int_{0}^{p}dqD_{p,q}(0)\hat{r}_{p}^{\dagger}\hat{r}_{q}^{\dagger}\vert\emptyset\rangle.
\end{eqnarray}
In terms of the basis wave function
\begin{eqnarray}
\langle x_{1},x_{2}\vert\hat{r}_{p}^{\dagger}\hat{r}_{q}^{\dagger}\vert0\rangle=\mathcal{N}_{rr}\left(e^{i\Delta_{p}x_{1}}e^{i\Delta_{q}x_{2}}+x_{1}\leftrightarrow x_{2}\right)
\end{eqnarray}
with $\mathcal{N}_{rr}=1/(2\sqrt{2}\pi)$, the wave function in position space of the initial state is
\begin{eqnarray}
\langle x_{1},x_{2}\vert\psi(0)\rangle & = & -4\pi^{2}\mathcal{N}_{rr} G_{2}e^{(i\delta_{1}+\epsilon)(x_{1}+l_{1})}e^{(i\delta_{2}+\epsilon)(x_{2}+l_{2})}\nonumber \\
 &  & \times\theta(-x_{1}-l_{1})\theta(-x_{2}-l_{2})+(x_{1}\leftrightarrow x_{2}).
\label{wafuncscaini}
\end{eqnarray}

By introducing the center-of-mass coordinate $x_{c}=(x_{1}+x_{2})/2$,
relative coordinate $x=x_{1}-x_{2}$, total center-detuning $E=\delta_{1}+\delta_{2}$, and relative center detuning $\delta=(\delta_{1}-\delta_{2})/2$, the wave
function (\ref{wafuncscaini}) becomes
\begin{eqnarray}
\langle x_{1},x_{2}\vert\psi(0)\rangle & = & -4\pi^{2}\mathcal{N}_{rr} G_{2}\left[e^{i\delta_{1}(x_{c}+x/2+l_{1})}e^{\epsilon(x_{c}+x/2+l_{1})}\right.\nonumber \\
 &  & \left.\times e^{i\delta_{2}(x_{c}-x/2+l_{2})}e^{\epsilon(x_{c}-x/2+l_{2})}\right.\nonumber \\
 &  & \times\left.\theta(-x_{c}+x/2-l_{2})\theta(-x_{c}-x/2-l_{1})\right.\nonumber \\
 &  & \left.+(x\leftrightarrow-x)\right].
\end{eqnarray}
For the special case of $l_{2}=l_{1}$, the wave function reduces to
\begin{eqnarray}
\langle x_{1},x_{2}\vert\psi(0)\rangle & = & -8\pi^{2}\mathcal{N}_{rr}G_{2}e^{(iE+2\epsilon)(x_{c}+l_{1})}\nonumber \\
 &  & \times\cos(\delta x)\theta(-x_{c}-l_{1}-|x|/2).
\end{eqnarray}

After being scattered by the two Rydberg atoms, the state of the two photons in the long-time limit can be expressed as
\begin{eqnarray}
\vert\psi(t\rightarrow\infty)\rangle & = & |\psi_{rr}\rangle+|\psi_{rl}\rangle+|\psi_{lr}\rangle+|\psi_{ll}\rangle,\label{eq:output}
\end{eqnarray}
where
\begin{eqnarray}
|\psi_{rr}\rangle & = & \int_{0}^{\infty}\int_{0}^{\infty}dpdqD_{p,q}^{rr}e^{-i(\Delta_{p}+\Delta_{q})t}\hat{r}_{p}^{\dagger}\hat{r}_{q}^{\dagger}\vert\emptyset\rangle,\nonumber \\
|\psi_{rl}\rangle & = & \int_{0}^{\infty}\int_{0}^{\infty}dpdqD_{p,q}^{rl}e^{-i(\Delta_{p}+\Delta_{q})t}\hat{r}_{p}^{\dagger}\hat{l}_{q}^{\dagger}\vert\emptyset\rangle,\nonumber \\
|\psi_{lr}\rangle & = & \int_{0}^{\infty}\int_{0}^{\infty}dpdqD_{p,q}^{lr}e^{-i(\Delta_{p}+\Delta_{q})t}\hat{l}_{p}^{\dagger}\hat{r}_{q}^{\dagger}\vert\emptyset\rangle,\nonumber \\
|\psi_{ll}\rangle & = & \int_{0}^{\infty}\int_{0}^{\infty}dpdqD_{p,q}^{ll}e^{-i(\Delta_{p}+\Delta_{q})t}\hat{l}_{p}^{\dagger}\hat{l}_{q}^{\dagger}\vert\emptyset\rangle,\label{eq:output_k}
\end{eqnarray}
with
\begin{eqnarray}
D_{p,q}^{rr} & = & \frac{1}{2}(T_{p}T_{q}D_{p,q}(0)+J_{p,q}/4),\nonumber \\
D_{p,q}^{rl} & = & \frac{1}{2}(T_{p}R_{q}D_{p,q}(0)+J_{p,q}/4),\nonumber \\
D_{p,q}^{lr} & = & \frac{1}{2}(R_{p}T_{q}D_{p,q}(0)+J_{p,q}/4),\nonumber \\
D_{p,q}^{ll} & = & \frac{1}{2}(R_{p}R_{q}D_{p,q}(0)+J_{p,q}/4).
\end{eqnarray}
Here, $D_{p,q}^{rr}$ and $D_{p,q}^{ll}$ are, respectively, the two-photon
transmission and reflection amplitudes. In addition,
$D_{p,q}^{rl}$ ($D_{p,q}^{lr}$) relates to the process where the
photon with wave number $p$ ($q$) is transmitted into the
right-propagation mode and the photon with wave number $q$ ($p$) is
reflected into the left-propagation mode.

\subsection{Two-photon correlation in position variables}

To characterize the photon correlation, we will consider the two-photon transmission and reflection cases. For the two-photon
transmission, the output state of the two right-going photons is
\begin{eqnarray}
\langle x_{1},x_{2}|\psi_{rr}\rangle & = & -4\pi^{2}\mathcal{N}_{rr}G_{2}e^{i(E-2i\epsilon)(x_{c}-t)}\nonumber \\
 &  & \times e^{i(E/2-i\epsilon)(l_{1}+l_{2})}e^{i\delta(l_{1}-l_{2})}\Phi_{rr}(x)\label{eq:psi_rr}
\end{eqnarray}
with
\begin{eqnarray}
\Phi_{rr}(x) & = & T_{\delta_{1}-i\epsilon}T_{\delta_{2}-i\epsilon}[e^{i\delta x}\theta(-x_{c}-x/2+t-l_{1})\nonumber \\
 &  & \times\theta(-x_{c}+x/2+t-l_{2})+x\leftrightarrow-x]\nonumber \\
 &  & -R_{\delta_{1}-i\epsilon}R_{\delta_{2}-i\epsilon}\frac{E-2i\epsilon-2\xi}{E-2i\epsilon-\xi+i\gamma}\nonumber \\
 &  & \times e^{i(E/2-i\epsilon+i\gamma)|x|}\theta(-x_{c}+t-|x|/2-l_{1}).
\end{eqnarray}
When $l_{2}=l_{1}$, the output state~(\ref{eq:psi_rr}) reduces to
\begin{eqnarray}
\langle x_{1},x_{2}|\psi_{rr}\rangle & = & -8\pi^{2}\mathcal{N}_{rr}G_{2}e^{(iE+2\epsilon)(x_{c}-t+l_{1})}\nonumber \\
 &  & \times\theta(-x_{c}+t-l_{1}-|x|/2)\phi_{rr}(x),\label{eq:psiR1}
\end{eqnarray}
 where
\begin{eqnarray}
\phi_{rr}(x) & = & T_{\delta_{1}-i\epsilon}T_{\delta_{2}-i\epsilon}\cos(\delta x)\nonumber \\
 &  & -\frac{1}{2}R_{\delta_{1}-i\epsilon}R_{\delta_{2}-i\epsilon}\frac{E-2\xi-2i\epsilon}{E-\xi-2i\epsilon+i\gamma}e^{i(E/2-i\epsilon+i\gamma)|x|},\nonumber \\
\label{eq:psiR2}
\end{eqnarray}
which satisfies $\phi_{rr}(-x)=\phi_{rr}(x)$. In the derivation
of Eq.~(\ref{eq:psi_rr}), we have used the condition $\gamma\gg\epsilon$.

We note that Eq.~(\ref{eq:psiR1}) is a product of the center-of-mass
wave function $\exp[(iE+2\epsilon)(x_{c}-t+l_{1})]$ and the relative
wave function $\phi_{rr}(x)$ in the region defined by the step function
$\theta(-x_{c}+t-l_{1}-|x|/2)$. As $|\phi_{rr}(x)|^{2}$ is proportional
to the joint probability for two photons with a separation $x$, it
could be used to characterize the spatial statistics of the two photons.
For example, the peak and dip feature of $|\phi_{rr}(x)|^{2}$ around zero distance $x=0$,
implies photon bunching and antibunching, respectively. In Eq.~(\ref{eq:psiR2}),
the first term of $\phi_{rr}(x)$ describes an independent two-photon
transmission process, while the second term is a two-photon correlation
induced by scattering. Physically, the present system has two resonant
scattering conditions: single- and two-photon resonances. When the
frequency center of a single-photon wave packet matches the energy separation of
a single atom, i.e. $\delta_{1}=\delta_{2}=0$ (or
$E=0$ and $\delta=0$), the single photon will resonantly excite
the atom. We call this as the single-photon resonance condition. On
the other hand, when the total center detuning of the two photons equals to
the energy shift due to the Rydberg coupling, i.e., $\delta_{1}+\delta_{2}=\xi$
(or $E=\xi$), the two photons can resonantly excite the two coupled
atoms, even the single-photon process could be off-resonant. This
regime is called two-photon resonance.
\begin{figure}
\includegraphics[bb=37 13 538 495,width=3.3in]{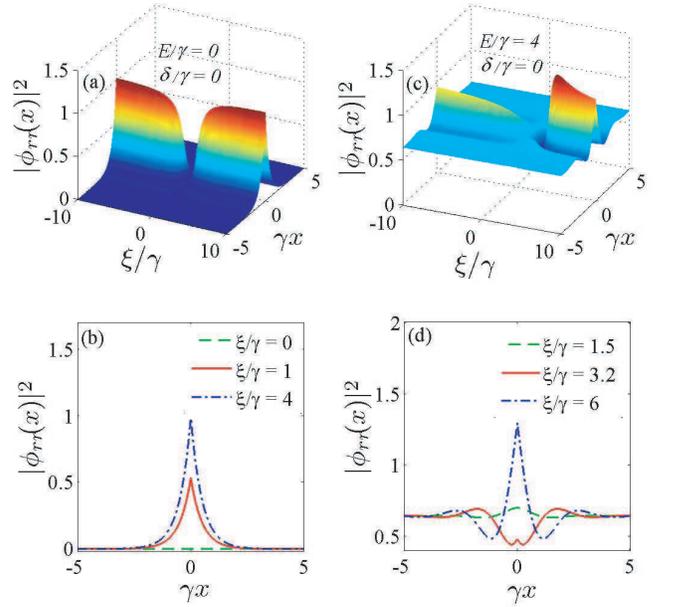}
\caption{(Color online) Plots of $|\phi_{rr}(x)|^{2}$ vs the scaled parameters
$\xi/\gamma$ and $\gamma x$ when (a) $E/\gamma=0$, $\delta=0$
and (b) $E/\gamma=4$, $\delta=0$. (c) and (d) show, respectively,
the curves from (a) and (b) when the parameter $\xi/\gamma$ takes
several certain values. Here, we consider the near monochromatic limit
$\epsilon/\gamma=0.01$.}
\label{twophotran}
\end{figure}

In the single-photon resonance case, the independent two-photon transmission
process will be completely suppressed, and a pure photon-correlation
effect can be seen from $|\phi_{rr}(x)|^{2}$. In Fig.~\ref{twophotran}(a),
we plot $|\phi_{rr}(x)|^{2}$ as a function of $\xi/\gamma$ and $\gamma x$.
When $\xi=0$, there is no photon correlation {[}dashed line in Fig.~\ref{twophotran}(b){]}.
This result corresponds to the fluorescence-complete-vanishing phenomenon
found in Ref.~\cite{Rephaeli2011}. For a nonzero $\xi/\gamma$,
we can see clear evidence for photon bunching {[}Fig.~\ref{twophotran}(b){]}.
In particular, with the increasing of $\xi/\gamma$, $|\phi_{rr}(0)|^{2}$
increases gradually, and saturates when $\xi/\gamma\gg1$. This means
that the photon bunching becomes stronger for a larger $\xi/\gamma$
in the single-photon resonance regime.

In the case of single-photon off resonance (e.g., $E/\gamma=4$), we plot
$|\phi_{rr}(x)|^{2}$ vs $\xi/\gamma$ and $\gamma x$ in Figs.~\ref{twophotran}(c)
and~\ref{twophotran}(d). The curves exhibit photon bunching in most
regions of $\xi/\gamma$, but there is also some oscillation
pattern with respect to $\gamma x$ due to independent photon transmission.
However, around the two-photon resonance, i.e., $\xi=E$, there
is a clear evidence for the photon antibunching {[}Fig.~\ref{twophotran}(d){]}.
This interesting phenomenon of photon statistics transition from bunching to antibunching is induced
by the two-photon resonance.

Similarly, using the basis wave function
\begin{eqnarray}
\langle x_{1},x_{2}\vert\hat{l}_{p}^{\dagger}\hat{l}_{q}^{\dagger}\vert0\rangle=\mathcal{N}_{ll}\left(e^{-i\Delta_{p}x_{1}}e^{-i\Delta_{q}x_{2}}+x_{1}\leftrightarrow x_{2}\right)
\end{eqnarray}
with $\mathcal{N}_{ll}=1/(2\sqrt{2}\pi)$, the output state of the two left-going photons
in the long-time limit $t\rightarrow\infty$ is
\begin{eqnarray}
\langle x_{1},x_{2}|\psi_{ll}\rangle & = & -4\pi^{2}\mathcal{N}_{ll}G_{2}e^{i(E/2-i\epsilon)(l_{1}+l_{2})}e^{i\delta(l_{1}-l_{2})}\nonumber \\
 &  & \times e^{i(E-2i\epsilon)(-x_{c}-t)}\Phi_{ll}(x)\label{eq:psi_ll}
\end{eqnarray}
 with
\begin{eqnarray}
\Phi_{ll}(x) & \equiv & R_{\delta_{1}-i\epsilon}R_{\delta_{2}-i\epsilon}[e^{-i\delta x}\theta(x_{c}+x/2+t-l_{1})\nonumber \\
 &  & \times\theta(x_{c}-x/2+t-l_{2})+x\leftrightarrow-x]\nonumber \\
 &  & -R_{\delta_{1}-i\epsilon}R_{\delta_{2}-i\epsilon}\frac{E-2i\epsilon-2\xi}{E-2i\epsilon-\xi+i\gamma}\nonumber \\
 &  & \times e^{i(E/2-i\epsilon+i\gamma)|x|}\theta(x_{c}+t-|x|/2-l_{1}).
\end{eqnarray}
 When $l_{1}=l_{2}$, the output state of two reflected photons becomes
\begin{eqnarray}
\langle x_{1},x_{2}|\psi_{ll}\rangle & = & -8\pi^{2}\mathcal{N}_{ll}G_{2}e^{(iE+2\epsilon)(-x_{c}-t+l_{1})}\nonumber \\
 &  & \times\theta(x_{c}+t-l_{1}-|x|/2)\phi_{ll}(x)
\end{eqnarray}
 with
\begin{eqnarray}
\phi_{ll}(x) & = & R_{\delta_{1}-i\epsilon}R_{\delta_{2}-i\epsilon}\left[\cos(\delta x)\right.\nonumber \\
 &  & \left.-\frac{1}{2}\frac{E-2i\epsilon-2\xi}{E-2i\epsilon-\xi+i\gamma}e^{i(\frac{E}{2}-i\epsilon+i\gamma)|x|}\right].
\end{eqnarray}
\begin{figure}
\includegraphics[bb=37 13 538 495,width=3.3in]{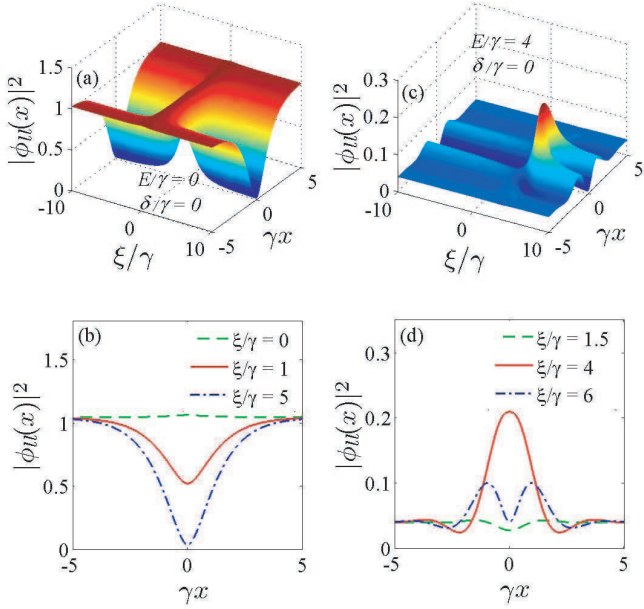}
\caption{(Color online) Plots of $|\phi_{ll}(x)|^{2}$ vs $\xi/\gamma$ and
$\gamma x$ when (a) $E/\gamma=0$, $\delta=0$ and (b) $E/\gamma=4$,
$\delta=0$. (c) and (d) are, respectively, the curves from (a) and
(b) when the parameter $\xi/\gamma$ takes several certain values.
Here, we take $\epsilon/\gamma=0.01$.}
\label{twophoref}
\end{figure}

In Figs.~\ref{twophoref}(a) and~\ref{twophoref}(b), we plot $|\phi_{ll}(x)|^{2}$
as a function of $\xi/\gamma$ and $\gamma x$ at the single-photon
resonance $E/\gamma=0$ and $\delta/\gamma=0$. Similar to the
two-photon transmission, when $\xi=0$, there is no photon correlation
{[}dash line in Fig.~\ref{twophoref}(b){]}. For nonzero $\xi$,
we can see evident photon antibunching [Fig.~\ref{twophoref}(b)]. With
the increasing of $\xi/\gamma$, the $|\phi_{ll}(0)|^{2}$ decreases
gradually, and eventually approaches zero when $\xi/\gamma\gg0$.

On the other hand, $|\phi_{ll}(x)|^{2}$ in the case of single-photon off resonance ($E/\gamma=4$)
is shown in Figs.~\ref{twophoref}(c)
and~\ref{twophoref}(d). Though there exists some oscillation with
respect to $\gamma x$, we can still see photon antibunching in most
region of the parameter $\xi/\gamma$. In addition, it is similar to the
two-photon transmission in the sense that there also exists the photon statistics
transition induced by the two-photon resonance. Around $\xi=E$, we
see clear evidence for photon bunching [Fig.~\ref{twophoref}(d)].
We find that this bunching peak is exactly at $\xi=E$ for the reflected
photons. This transition of photon statistics can be used to detect the form of
the interaction between atoms.

\subsection{Second-order correlation function}

We can also present a quantitative description of the statistics of the right- and left-going photons
using the second-order correlation function $g^{(2)}$. In particular, we are only concerned with the two-photon
reflection and transmission because the photon statistics in these two cases makes sense.
In terms of the coordinates of the two photons, the component of the two-photon state in Eq.~(\ref{eq:output}) can be reexpressed as
\begin{eqnarray}
|\psi_{ss'}\rangle & = & \frac{1}{\sqrt{2}}\int\int dx_{1}dx_{2}\langle x_{1},x_{2}|\psi_{ss'}\rangle
\hat{\psi}_{s}^{\dagger}(x_{1})\hat{\psi}_{s'}^{\dagger}(x_{2})|\emptyset\rangle\nonumber \\
\label{eq:output_x}
\end{eqnarray}
for $s,s'=r,l$, where the field operators satisfy the bosonic commutation relation $[\hat{\psi}_{s}(x_{1}),\hat{\psi}_{s}^{\dagger}(x_{2})]=\delta(x_{1}-x_{2})$.
For state~(\ref{eq:output_x}), the second-order correlation function is
\begin{eqnarray}
g_{s}^{(2)}(\tau)&=&\frac{G_{s}^{(2)}(x_{1},\tau)}{G_{s}^{(1)}(x_{1})G_{s}^{(1)}(x_{1}+\tau)},\hspace{0.5 cm} s=r,l,\label{eq:g2_def}
\end{eqnarray}
where
\begin{eqnarray}
G_{s}^{(1)}(x)&=&\langle \bar{\psi}_{ss}|\hat{\psi}_{s}^{\dagger}(x)\hat{\psi}_{s}(x)|\bar{\psi}_{ss}\rangle,\nonumber\\
G_{s}^{(2)}(x_{1},\tau)&=&\langle \bar{\psi}_{ss}|\hat{\psi}_{s}^{\dagger}(x_{1})\hat{\psi}_{s}^{\dagger}(x_{1}+\tau)\hat{\psi}
_{s}(x_{1}+\tau)\hat{\psi}_{s}(x_{1})|\bar{\psi}_{ss}\rangle,  \nonumber \\
\end{eqnarray}
with $|\bar{\psi}_{ss}\rangle=|\psi_{ss}\rangle/\sqrt{\langle
\psi_{ss}|\psi_{ss}\rangle }$. Then, combination of Eq.~(\ref{eq:output_x}) with Eq.~(\ref{eq:g2_def}) yields
\begin{eqnarray}
g_{s}^{(2)}(\tau)&=&\frac{|\psi_{ss}(x_{1},x_{1}+\tau)|^{2}\int\int
dx_{1}dx_{2}|\psi_{ss}(x_{1},x_{2})|^{2}}{2\int
dx_{2}|\psi_{ss}(x_{1},x_{2})|^{2}\int
dx_{2}|\psi_{ss}(x_{1}+\tau,x_{2})|^{2}},\notag \\\label{eq:g2}
\end{eqnarray}
for $s=r,l$.

For the two transmitted photons, we introduce a new frame of reference as
$x_{1}=x_{1}^{\prime}+t$ and $x_{2}=x_{2}^{\prime}+t$.
Correspondingly, the center-of-mass and relative coordinates become
$x_{c}^{\prime}=(x_{1}^{\prime}+x_{2}^{\prime})/2$ and
$x^{\prime}=x_{1}^{\prime}-x_{2}^{\prime}$, which have the same form as those in the old frame of reference of $x$.
In the new frame of reference, the state of the two right-going photons is
\begin{eqnarray}
\langle x_{1}^{\prime },x_{2}^{\prime}|\psi_{rr}\rangle&=&-8\pi^{2}
\mathcal{N}_{rr}G_{2}e^{(iE+2\epsilon)(x_{c}^{\prime }+l_{1})}\nonumber\\
& & \times\theta(-x_{c}^{\prime }-l_{1}-|x^{\prime}|/2)\phi_{rr}(x^{\prime}),
\end{eqnarray}
which is independent of the time $t$. In the following, we will restrict
our calculation in the new frame of reference, and omit the
superscript ``$^{\prime }$'' for simplicity. After some tedious
calculations, we get
\begin{equation}
g_{r}^{(2)}(\tau)=F_{r}(x_{1},\tau)|\phi_{rr}(\tau)|^{2},\label{eq:g2R}
\end{equation}
where
\begin{eqnarray}
F_{r}(x_{1},\tau)&=&M_{r}^{-1}\theta(-x_{1}-l_{1}-\tau)\theta(-x_{1}-l_{1})\nonumber\\
&&\times\int_{0}^{+\infty}e^{-2\epsilon x}|\phi_{rr}(x)|^{2}dx ,
\end{eqnarray}
with
\begin{eqnarray}
M_{r}&=&4\epsilon\theta(-x_{1}-l_{1}-\tau)\theta(-x_{1}-l_{1})e^{4\epsilon(x_{1}+l_{1}+\frac{\tau}{2})}\nonumber\\
&&\times\int_{x_{1}+l_{1}}^{\infty}dxe^{-2\epsilon x}|\phi_{rr}(x)|^{2}\int_{x_{1}+\tau+l_{1}}^{\infty}dxe^{-2\epsilon x}|\phi_{rr}(x)|^{2}.\nonumber\\
\end{eqnarray}

Similarly, for the two reflected photons, we also introduce a new frame of reference as $x_{1}=x_{1}^{\prime }-t$ and $x_{2}=x_{2}^{\prime }-t$,
then the state for the two reflected photons becomes
\begin{eqnarray}
\langle x_{1}^{\prime },x_{2}^{\prime}|\psi_{ll}\rangle&=&-8\pi^{2}\mathcal{N}
_{ll}G_{2}e^{(iE+2\epsilon)(-x_{c}^{\prime }+l_{1})}\nonumber\\
& & \times\theta(x_{c}^{\prime }-l_{1}-|x^{\prime}|/2)\phi_{ll}(x^{\prime}),
\end{eqnarray}
and the correlation function $g_{ll}^{(2)}(\tau)$ reads
\begin{equation}
g_{l}^{(2)}(\tau)=F_{l}(x_{1},\tau)|\phi_{ll}(\tau)|^{2},\label{eq:g2L}
\end{equation}
where
\begin{eqnarray}
F_{l}(x_{1},\tau)&=&M^{-1}_{l}\theta(x_{1}-l_{1}+\tau)\theta(x_{1}-l_{1})\notag \\
&&\times\int_{0}^{+\infty}e^{-2\epsilon x}|\phi_{ll}(x)|^{2}dx ,
\end{eqnarray}
with
\begin{eqnarray}
M_{l}&=&4\epsilon\theta(x_{1}-l_{1}+\tau)\theta(x_{1}-l_{1})e^{-4\epsilon(x_{1}-l_{1}+\frac{\tau}{2})}\nonumber\\
&&\times\int_{l_{1}-x_{1}}^{+\infty}e^{-2\epsilon x}|\phi_{ll}(x)|^{2}dx\int_{l_{1}-(x_{1}+\tau)}^{+\infty}e^{-2\epsilon x}|\phi_{ll}(x)|^{2}dx.\nonumber\\
\end{eqnarray}
\begin{figure}
\includegraphics[bb=37 10 538 495, width=3.3 in]{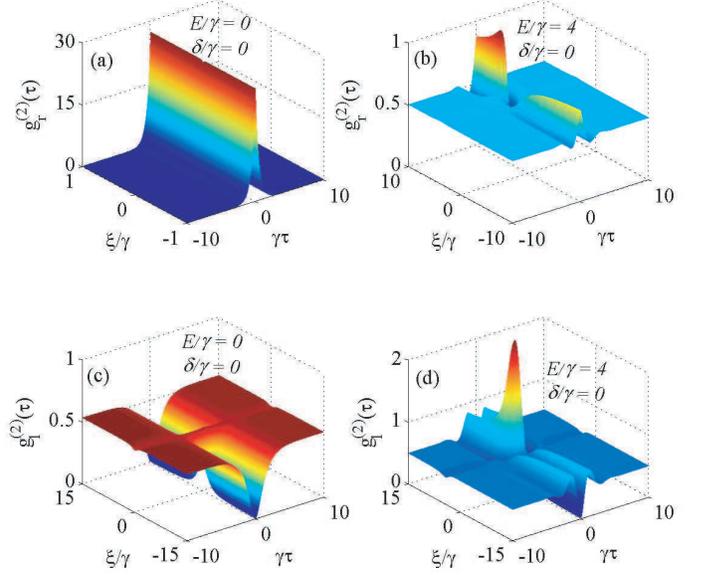}
\caption{(Color online) The second-order correlation function $g_{r}^{(2)}(\tau)$ and $g_{l}^{(2)}(\tau)$ vs $\xi/\gamma$ and
$\gamma \tau$ when (a),(c) $E/\gamma=0$, $\delta=0$ and (b),(d) $E/\gamma=4$,
$\delta=0$.}\label{g2}
\end{figure}

In Fig.~\ref{g2}, we illustrate the second-order correlation function $g_{r}^{(2)}(\tau)$ and $g_{l}^{(2)}(\tau)$ as
a function of the scaled parameters $\xi/\gamma$ and $\gamma\tau$. Here, the parameters are chosen to be the same as those
in Figs.~\ref{twophotran} and~\ref{twophoref} for comparison.
At the single-photon resonance, i.e., $E/\gamma=0$ and $\delta/\gamma=0$, we can see that, when $\xi/\gamma\neq0$, $g_{r}^{(2)}(0)>g_{r}^{(2)}(\tau)$  [Fig.~\ref{g2}(a)]
and $g_{l}^{(2)}(0)<g_{l}^{(2)}(\tau)$ [Fig.~\ref{g2}(c)], which represent photon bunching in transmission and photon antibunching in reflection, respectively.
Interestingly, the $g_{l}^{(2)}(0)$ approaches zero when $\xi/\gamma\gg1$, which means that the photon blockade phenomenon takes place in the reflection [Fig.~\ref{g2}(c)].
When $\xi/\gamma=0$, the two photons are reflected completely and independently. At the point $\xi/\gamma=0$, $g_{l}^{(2)}(\tau)=1/2$ (for Fock state $|2\rangle$) and $g_{r}^{(2)}(\tau)$ does not make sense.

On the other hand, Figs.~\ref{g2}(b) and~\ref{g2}(d) show that the two-photon resonance ($\xi=E$) will induce
the photon statistics transition between bunching and antibunching. For the two-photon transmission, we see $g_{r}^{(2)}(0)>g_{r}^{(2)}(\tau)$ (bunching) in most
region of $\xi/\gamma$ and $g_{r}^{(2)}(0)<g_{r}^{(2)}(\tau)$ (antibunching) around the two-photon resonance point $\xi=E$ [Fig.~\ref{g2}(b)]. For comparison,
the result for the two-photon reflection is shown in Fig.~\ref{g2}(d). We can see $g_{l}^{(2)}(0)>g_{l}^{(2)}(\tau)$ (bunching) around $\xi=E$ and $g_{l}^{(2)}(0)<g_{l}^{(2)}(\tau)$ (antibunching)
in other regions. These results are consistent with our analysis on photon correlation given in Sec.~\ref{sec-twophosca}(c).

\section{Conclusion\label{sec-conclu}}

To conclude, we have studied the transport of photonic wave packets in a 1D waveguide, controlled by the Rydberg
interaction between two atoms. We have found that the quantum statistical properties of
the scattered photons can be predicted from the relative wave function of the two photons as well as the second-order
correlation function. With detailed calculations about the relative wave functions, we have also pointed out how to control
the statistics of the scattered photons in the confined system. On one hand, for a certain Rydberg interaction
strength, by adjusting the photon-atom detuning, it is possible to
control the photon-statistics transition between bunching and antibunching. On the other hand, for a certain total center detuning, we can regulate the Rydberg interaction
strength by varying the distance between the two atoms or select two different energy levels of the Rydberg atoms.
We can use the change of the quantum statistic properties to detect the detail fashion
of the Rydberg interaction between the two atoms.

\begin{acknowledgments}
C. P. Sun was supported by National Natural Science Foundation of China
under Grants No. 11121403, No. 10935010, No. 11074261, and National 973 program
(Grants No.~2012CB922104). J. Q. Liao is
supported by Japan Society for the Promotion of Science (JSPS) Foreign Postdoctoral Fellowship No. P12503.
\end{acknowledgments}

\appendix
\section{SOLUTION OF EQ.~(\protect\ref{eq:motion_eq}) BY THE LAPLACE TRANSFORM}

In this appendix, we present a detailed derivation of Eq.~(\ref{eq:Dpq}) by employing the Laplace transform.
Under the initial condition~(\ref{eq:Ini}), the equation of motion~(\ref{eq:motion_eq}) becomes
\begin{eqnarray}
s\tilde{A}(s) & = & -i\xi\tilde{A}(s)-ig\int_{0}^{\infty}dk[\tilde{B}_{k}(s)+\tilde{C}_{k}(s)],\nonumber \\
s\tilde{B}_{k}(s) & = & -i\Delta_{k}\tilde{B}_{k}(s)-ig\tilde{A}(s)-ig\int_{0}^{\infty}dp\tilde{D}_{p,k}(s),\nonumber \\
s\tilde{C}_{k}(s) & = & -i\Delta_{k}\tilde{C}_{k}(s)-ig\tilde{A}(s)-ig\int_{0}^{\infty}dp\tilde{D}_{p,k}(s),\nonumber \\
s\tilde{D}_{p,q}(s) & = & D_{p,q}(0)-i(\Delta_{p}+\Delta_{q})\tilde{D}_{p,q}(s)\nonumber \\
 &  & -ig[\tilde{B}_{p}(s)+\tilde{B}_{q}(s)+\tilde{C}_{p}(s)+\tilde{C}_{q}(s)].\label{eq:Le4}
\end{eqnarray}
By eliminating other variables, we obtain the following equation for the variable $\tilde{B}_{k}(s)$:
\begin{eqnarray}
 &  & [\Delta_{k}-i(s+\gamma)]\tilde{B}_{k}(s)\nonumber \\
 & = & 2g^{2}\int_{-\infty}^{\infty}d\Delta_{p}\left(\frac{1}{\xi-is}+\frac{1}{\Delta_{p}+\Delta_{k}-is}\right)\tilde{B}_{p}(s)\nonumber \\
 &  & +ig\int_{-\infty}^{\infty}\frac{D_{p,k}(0)}{\Delta_{p}+\Delta_{k}-is}d\Delta_{p}.\label{eq:B}
\end{eqnarray}
The solution of $\tilde{B}_{k}(s)$ can be obtained as
\begin{eqnarray}
\tilde{B}_{k}(s) & = & \frac{2\pi gG_{2}}{\Delta_{k}-i(s+\gamma)}[\widetilde{f}_{1}+2e^{-sl_{1}}\gamma(\widetilde{f}_{2}+\widetilde{f}_{3})]\nonumber \\
 &  & \times\Theta(l_{1})\Theta(l_{2})\Theta(l_{1}-l_{2}),
\end{eqnarray}
with
\begin{widetext}
\begin{eqnarray}
\widetilde{f}_{1} & = & \frac{e^{-sl_{1}}e^{-i\left(l_{1}-l_{2}\right)\Delta_{k}}}{\left(\Delta_{k}-\delta_{2}+i\epsilon\right)\left(\Delta_{k}+\delta_{1}-i\left(s+\epsilon\right)\right)}
+\frac{e^{-sl_{2}}e^{i\left(l_{1}-l_{2}\right)\Delta_{k}}}{\left(\Delta_{k}-\delta_{1}+i\epsilon\right)\left(\Delta_{k}+\delta_{2}-i\left(s+\epsilon\right)\right)},\nonumber\\
\widetilde{f}_{2}&=&\frac{e^{-i\left(l_{1}-l_{2}\right)\Delta_{k}}}{\left(\Delta_{k}+i\gamma\right)
\left(\Delta_{k}-\delta_{2}+i\epsilon\right)\left(s+\epsilon+i\left(\Delta_{k}+\delta_{1}\right)\right)}
+\frac{e^{-\left(l_{1}-l_{2}\right)\gamma}}{\left(\gamma-\epsilon-i\delta_{2}\right)\left(s+\gamma+\epsilon+i\delta_{1}\right)}
\left(\frac{1}{-\gamma+i\Delta_{k}}+\frac{1}{s+\gamma+i\xi}\right),\nonumber \\
\widetilde{f}_{3} & = & \frac{e^{-\left(l_{1}-l_{2}\right)\left(i\delta_{2}+\epsilon\right)}}{s+i\delta_{1}+i\delta_{2}+2\epsilon}\left(
\frac{1}{\left(i\gamma+\delta_{2}-i\epsilon\right)\left(\delta_{2}-\Delta_{k}-i\epsilon\right)}
-\frac{1}{\left(s+\gamma+i\delta_{2}+\epsilon\right)\left(s+i\delta_{2}+i\Delta_{k}+\epsilon\right)}\right.\nonumber\\
&&\left.-\frac{1}{\left(\gamma-i\delta_{2}-\epsilon\right)\left(s+\gamma+i\xi\right)}-\frac{1}{\left(s+\gamma+i\delta_{2}+\epsilon\right)\left(s+\gamma+i\xi\right)}\right).
\end{eqnarray}
\end{widetext}
Then we have
\begin{equation}
\tilde{D}_{p,q}(s)=\frac{D_{p,q}(0)-2ig[\tilde{B}_{p}(s)+\tilde{B}_{q}(s)]}{s+i(\Delta_{p}+\Delta_{q})}.
\end{equation}
The transient solution of $D_{p,q}(t)$ can be obtained using the inverse Laplace transform.
For studying photon scattering, it is sufficient to get the long-time solution given in Eq.~(\ref{eq:Dpq}).

%\end{CJK*}
\end{document}